\documentclass[pre,a4paper,onecolumn,notitlepage,nofootinbib]{revtex4-1}
\usepackage{url,hyperref,lineno,microtype}
\usepackage[pdftex]{graphicx}

\begin{document}

\title[Revisiting nldFC]{Revisiting non-linear functional brain co-activations: directed, dynamic and delayed} 

\author{Ignacio Cifre}
\email{Corresponding author: ignaciocl@blanquerna.url.edu}
\address{Facultat de Psicologia, Ci\`encies de l'educaci\'o  i de l'Esport, Blanquerna, Universitat Ramon Llull, Barcelona, Spain}\affiliation{Center for Complex Systems \& Brain Sciences (CEMSC$^3$), Escuela de Ciencia y Tecnolog\'ia,  Universidad Nacional de San Mart\'{i}n, Buenos Aires, Argentina}

\author{Maria T. Miller Flores}
\address{Center for Complex Systems \& Brain Sciences (CEMSC$^3$), Escuela de Ciencia y Tecnolog\'ia,  Universidad Nacional de San Mart\'{i}n, Buenos Aires, Argentina}

\author{Jeremi K. Ochab}
\address{Institute of Theoretical Physics and Mark Kac Center for Complex Systems Research, Jagiellonian University, Krak\'ow, Poland}
\author{Dante R. Chialvo}
\address{Center for Complex Systems \& Brain Sciences (CEMSC$^3$), Escuela de Ciencia y Tecnolog\'ia,  Universidad Nacional de San Mart\'{i}n, Buenos Aires, Argentina}\affiliation{Consejo Nacional de Investigaciones Cient\'{i}ficas y Tecnol\'{o}gicas (CONICET), Buenos Aires, Argentina}
\date{\today}

\begin{abstract}
The center stage of neuro-imaging is currently occupied by studies of functional correlations between brain regions. These correlations define the brain functional networks, which are the most frequently used framework to represent and interpret a variety of experimental findings. In previous work we first demonstrated  that the relatively stronger BOLD activations contain most of the information relevant to understand functional connectivity and subsequent work  confirmed that a large compression of the original signals  can be obtained without significant loss of information. In this work we revisit  the correlation properties of these epochs to define a measure of nonlinear dynamic directed functional connectivity (\emph{nldFC}) across regions of interest.  
We show that the proposed metric provides at once, without extensive numerical complications, \emph{directed} information of the functional correlations, as well as a measure of ~\emph{temporal lags} across regions, overall offering a different perspective in the analysis of brain co-activation patterns. 
In this paper we provide for a proof of concept, based on replicating and completing existing results on an Autism database, to discuss the main features and advantages of the proposed strategy for the study of brain functional correlations. These results show new interpretations of the correlations found on this sample. 

\end{abstract}

\maketitle

\section{Introduction}

The large scale dynamics of the brain exhibits a plethora of  spatio-temporal patterns.  Since the first description of voxel-wise correlation networks~\cite{Eguiluz2005} there has been a continuous interest in developing better ways to derive  brain ``networks'' from fMRI time series data. Common to all is the  identification of functional ``nodes'' ( i.e., fMRI time series extracted from region of interest (ROI)), functional edges (i.e., the relatively stronger cross-correlations  which allows for the subsequent graph analysis.  An important methodological challenge has been always to define an adequate coarse graining of the brain imaging data  to  compress  thousands of the so-called BOLD (``blood oxygenated level dependent'') time series. The usual analysis aims at the identification of bursts of correlated activity across certain regions, which requires extensive computations, complicated in part by the humongous size of the data sets. 

In earlier work we proposed that the timing of the brief epochs of relatively stronger BOLD activations contain  a great deal of  functional connectivity (FC)  information~\cite{Tagliazucchi2011,Tagliazucchi2012}. The results of subsequent work~\cite{Tagliazucchi2016,Petridou2013,Allan2015,Li2014,Liu2013b,Liu2013,Chen2015,Jiang2014,Amico2014,Wu2013} seems to provide ample support to this idea, by confirming the functional relevance of such relatively large amplitude BOLD events under a variety of conditions.

The present work goes beyond the analysis of correlations between BOLD time series to explore and define a set of measures of the \emph{nonlinear directed dynamic functional correlation} across regions of interest (ROIs). The use of such measures, despite its simplicity, may help to expand at once the perspective of the usual FC paradigms, such as seed correlations maps and networks, into the realms of nonlinear time-dependent directed correlations.  


The paper is organized as follows: in the next section  we describe the essence of the method, starting with the basic procedure to define the BOLD triggered events followed by a description of the available correlation measures that allow a proper definition of the functional connectivity between the events, including definition of directionality and temporal lag of the events. Section 3 contains the analysis of a simple example as a  proof of concept of a healthy subjects fMRI data set,  followed by the replication and further analysis of a voxel-wize  published data set from Autism Syndrome in order to show the method features. The paper closes with a discussion of the advantages and limitations of the method and potential implications of the results. Derivations and further technical details are condensed on the Appendix.

\section{Methods}
\label{sec:meth}

The analysis to be discussed can use BOLD time series recorded indistinctly from either resting state conditions or during an experiment in which the subject is performing a given task (for space reasons we only show in the subsequent sections results on resting state conditions).  The most common approach to determine functional connectivity is to compute Pearson linear correlation between BOLD time series \cite{Finn2015,VandenHeuvel2010}. In contrast, the objective of the present analysis is to determine the relation between relatively large amplitude BOLD activations from a given pair of signals. In this section it will be discussed: \ref{sec:meth:event} how large amplitude events are selected given series of fMRI data; \ref{sec:meth:corr} correlations computed with the selected events; \ref{sec:meth:dir} how directionality is understood when working with events; and \ref{sec:meth:lags} how the dynamic connectivity, understood here as lags between time series, is computed.

\subsection{Definition of BOLD triggered events} \label{sec:meth:event}
\begin{figure}[h!]
\centering
\includegraphics[width=0.75\textwidth]{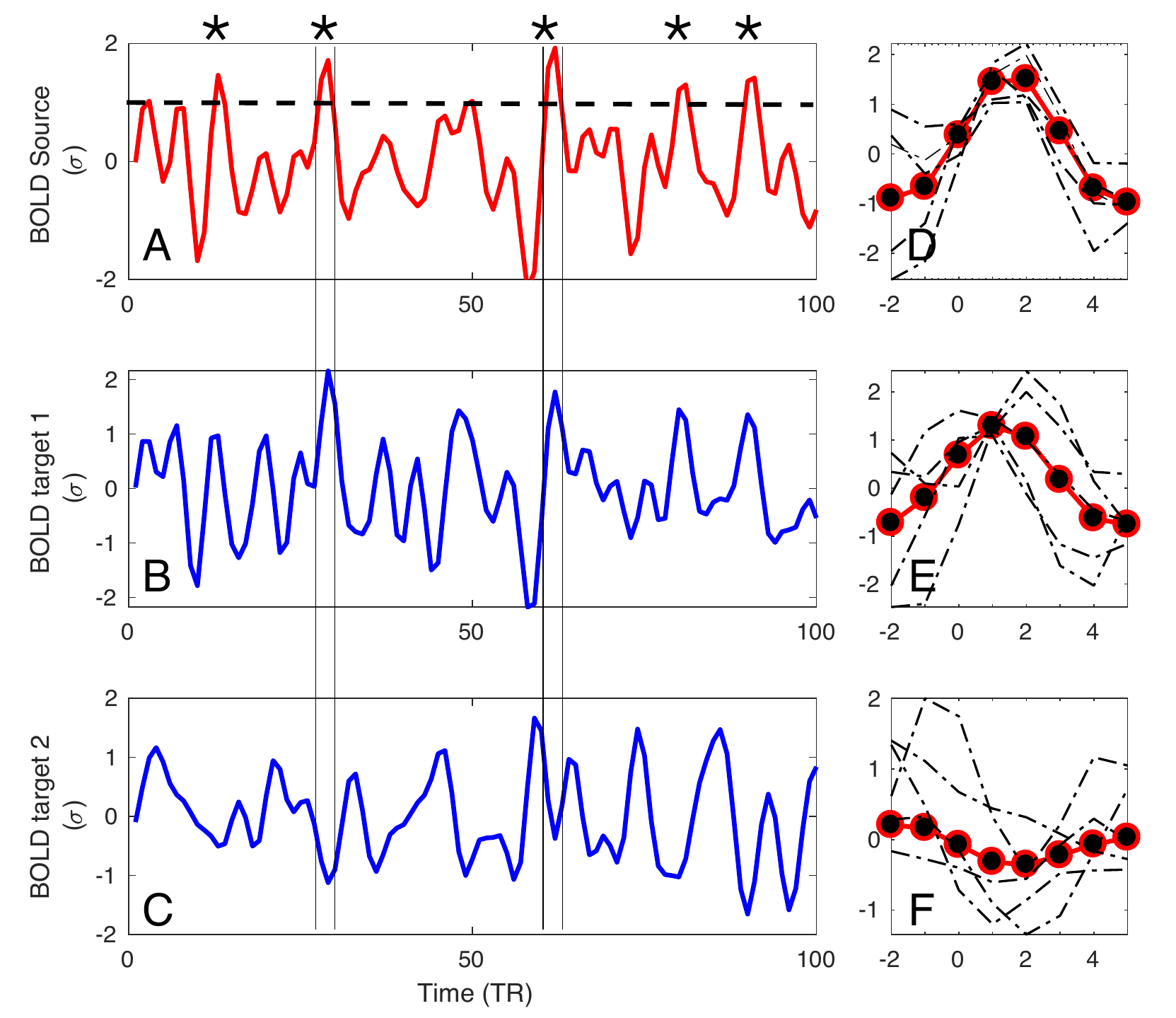}
\caption{Definition of the large amplitude events: For a given source region (panel A), BOLD triggered events (asterisks) are defined at the time at which the BOLD signal crosses an arbitrary threshold (here set to $1 \sigma$, denoted by the dashed line in panel A). To each source event correspond targets events which are defined by sampling at the same time intervals the BOLD signals at the regions of interest (as in the two examples in panels B and C denoted by vertical lines). Subsequently the extracted events can be averaged (Panels E-F), and eventually used to compute correlations, delays and directionality.}
\label{SelectEvents}
\end{figure}

First, each BOLD time series is z-scored (its mean is subtracted and it is divided by its standard deviation).
Next, a threshold for detecting strong activity is chosen, (typically the results remain unchanged when using a range of $1-2$ standard deviations)  and for each time series, the timing of each upward threshold crossing is determined (Fig.~\ref{SelectEvents}, panel A).
Note, that the number of threshold crossings depends on the auto-correlation of the BOLD signals (which stays in the range 0.6-0.85~\cite{Ochab2019}), and more generally on the exponent of the $1/f^{\alpha}$ frequency spectrum.
Empirically, for the threshold of $1 \sigma$, in a BOLD signal we find on average $8.5\pm2.8$ upward crossings per $4$ minutes of fMRI scan.

The timing is further used to define the \emph{seed} or \emph{source} events.
For a given seed voxel or region of interest (ROI) they consist of segments of BOLD time series starting typically $4-5 s$ before and ending $9-15 s$ after the crossing (which translates to $2-3$ TRs before and $4-7$ TRs after, with $TR=2.3$ in the data we are using as a proof of concept in this study). This time scale is chosen by the typical duration of these events, which in turn is dictated by the longest time scale of the haemodynamic response function ($\sim 10-15$ seconds).

Finally, for each seed event, the \emph{target} events are extracted from all the other BOLD time series at the exact same times as the seed, see Fig.~\ref{SelectEvents} B-C.
The average time courses of the events follow typically a smooth pattern, although they do exhibit variability, for both the seed (see Fig.~\ref{SelectEvents} D) and targets (see Fig.~\ref{SelectEvents} E-F).
If the interest of a given experiment is to define an average inter-relation measure between ROIs, then all the seed and target events might be averaged (as shown by red-and-black circles in Panels D-F of Fig.~\ref{SelectEvents}), for instance over the entire scan fMRI session.

\subsection{Correlations} \label{sec:meth:corr}
Once events are selected, a few  options of computing correlations are possible:
\begin{enumerate}
    \item $r_{P}(i,j)$ linear Pearson correlation between the whole time series $i$ and the whole time series $j$ (computed in section III were we perform a proof of concept). This option is not related to events, but in next section we will compute it to compare with next options,\label{corr:Pearson}
    \item $r_{E}^{(k)}(i,j)$ linear correlation between a $k$-th source event in time series $i$ and a respective target event in time series $j$ \label{corr:event_rBeta}. This option seems the most plausible when analyzing transient events, for instance localized tics on a motor disease, but we are not showing results here due to space limitations. 
    \item $\bar{r}_{E}(i,j)=1/K\sum_{k=1}^{K} r_{E}^{(k)}(i,j)$ average linear correlation between $K$ source events in time series $i$ and respective target events in time series $j$,\label{corr:mean_event_rBeta}
    \item $r_{C}(i,j)$ linear correlation between concatenated source events in time series $i$ and concatenated respective target event in time series $j$,\label{corr:concat_rBeta}
    \item $r_{E}(i,j)$ linear correlation between an average source event in time series $i$ and an average target event in time series $j$ (computed in section III were we perform a proof of concept).\label{corr:mean_rBeta}
\end{enumerate}
The choices, \ref{corr:mean_event_rBeta}-\ref{corr:concat_rBeta}, even we are not showing results on this paper, can be helpful in getting statistically less biased estimators of correlations.
In the paper we use \ref{corr:Pearson} and \ref{corr:mean_rBeta}.

\subsection{Directionality} \label{sec:meth:dir}
\begin{figure}
    \centering
    \includegraphics[width=\textwidth]{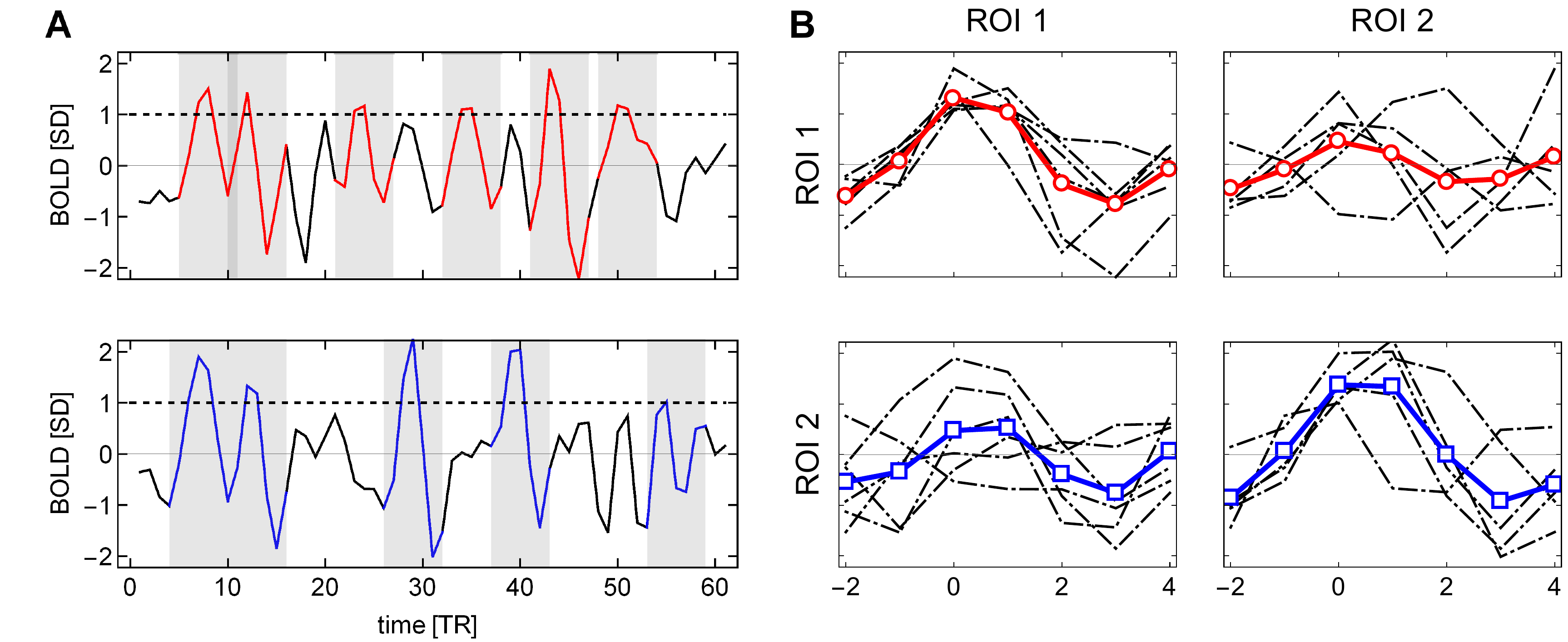}
    \caption{Example of asymmetry in source events of two regions of interest.
    (A) The shaded areas indicate the location of source events (BOLD activity around threshold crossing). The source events of ROI 2 may appear at different times than source events of ROI 1 (e.g., around $TR=30$ and $TR=40$).
    (B) Individual events (in gray) and their averages (in red for ROI 1 and blue for ROI 2); source events are shown in the diagonal subplots, and target events in the off-diagonal ones.
    Different sets of source events for each ROI give rise to asymmetry in the correlations between any two regions.}
    \label{AsymmetricEvents}
\end{figure}

Given two regions of interest $i$ and $j$, the linear Pearson correlation between their BOLD time series by definition is symmetric, i.e., $r_{P}(i,j)=r_{P}(j,i)$. It is not the case, if the correlations are computed using events. Then, the distinction between source and target becomes relevant, as shown in Fig.~\ref{AsymmetricEvents} A.
The shaded areas in the plots mark the positions of source events of each of the two relatively strongly correlated ROIs. Visibly, the first two events are common for both time series, but for instance the BOLD activations around $TR=30$ and $TR=40$ are source events for ROI 2 but not for ROI 1.

Consequently, the set of events over which one computes correlations when ROI 1 is the source is different from when ROI 2 is the source, which is shown in Fig.~\ref{AsymmetricEvents} B.
In that panel, the first row shows source events of ROI 1 and target events in ROI 1 in response to ROI 2 activations; the second row shows target events in ROI 2 in response to ROI 1 activations and source events of ROI 2. So even though the BOLD series of both regions are highly correlated, the source and target events are different, and hence the event correlation is not symmetric $r_{E}(i,j)\neq r_{E}(j,i)$.

The asymmetry in correlations might indicate directionality of co-activations between regions.
We can compute and assess a global correlation asymmetry of the whole brain
\begin{equation}
A=\sum_{i,j}\left( r_{E}(i,j)-r_{E}(j,i)\right),
\end{equation}
or similarly asymmetry of each ROI, or of each pair of time series $i$ and $j$.

Alternatively, in the spirit of analysis of point processes~\cite{Tagliazucchi2012,Tagliazucchi2016,cifre2017few}, we can use another proxy for directionality, based on the relative number of events occurring simultaneously in two regions.
For instance, in Fig.~\ref{AsymmetricEvents}, there are 2 out of 6 source events in ROI 1 that are also triggers (i.e., above threshold) in ROI 2, and 2 out of 5 in ROI 2 that are also triggers in ROI 1.
This approach takes into account event amplitudes, which to a large extent could be also achieved by computing covariance instead of correlation of events.
Below we call such ratio event directionality.

 \begin{figure} [h!]
 \centering
 \includegraphics[width=.5\textwidth]{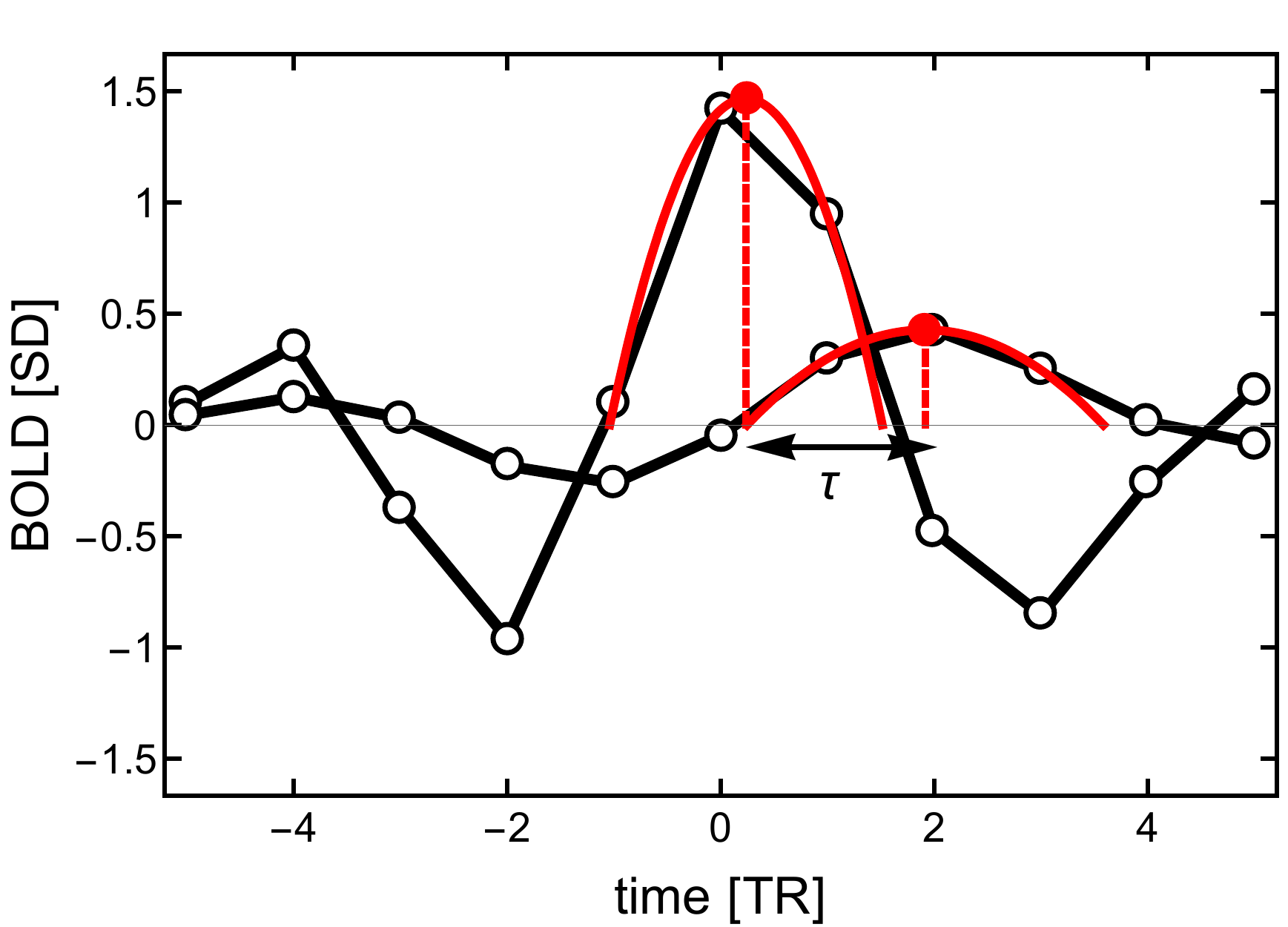}
 \caption{Definition of a lag $\tau$ between two events with a better resolution than sampling. The source event of one signal is centered around time $TR=0$, and the closest peak of target signal is at time $TR=2$. To obtain better precision of the delay, parabolas are fitted to three points in each peak. The time between the peaks of the parabolas defines the lag $\tau$.}
 \label{ComputationsExplained}
 \end{figure}

\subsection{Temporal lags}
\label{sec:meth:lags}
Recent studies \cite{Mitra2014,Mitra2015,Mitra2015a,Mitra2016,Mitra2018} have reported that the latency structure of the fluctuations in the fMRI BOLD signal shows spontaneous activity propagating through and across the cortex on a timescale of about 1s. This information was obtained by conventional lagged cross-covariance between pairs of BOLD time series $x_i(t)$ and $x_j(t)$ extracted from regions $i$ and $j$:
\begin{equation}
\text{C}_{i,j}(\tau)=\frac{1}{T} \sum_{t=1}^{T}  x_i(t+\tau)  x_j(t)
\end{equation}
where  $\tau$ is the lag (in TRs).
The value of $\tau(i,j)$ at which $\text{C}_{i,j}(\tau)$ exhibits an extremum defines the delay between signals $x_i$ and $x_j$.
Performing a parabolic interpolation of the cross-covariance extremum allows to determine the temporal lags with a finer resolution~\cite{Mitra2014}.
By definition, for a long time series the time delay matrix $\tau(i,j)$ is anti-symmetric, i.e., $\tau(i,j)=-\tau(j,i)$. 
Information on the cross-covariance value and the lags can be used to determine the structure of entire spatiotemporal processes.

Here, instead of computing cross-covariance between entire time series,
we make use of the fact that the BOLD triggered events have a well defined timing.
Given a source time series $x_i(t)$ and a target time series $x_j(t)$,
we obtain a set of $k_i$ source events.
For each source event in $x_i(t)$ we find the closest peak in $x_j(t)$ irrespective of its size and whether it occurred before or after the source event.
We search for the peak within a window of $[-6,8]$ TRs from the source threshold crossing.
As shown in Fig.~\ref{ComputationsExplained}, to obtain a finer timing of both the source and target peak we also use a parabolic fit.
The lag $\tau(i,j)$ is then defined as a difference between the target and source parabola peaks.
As a technical side note, when getting a peak value at the left or right edge of the time window we do not perform the parabola peak estimation, which could have unbounded values, but we set the lag to $-6$ or $6$, respectively.

Since the sets of source (threshold crossing) events of $x_i(t)$ and $x_j(t)$ can be (and usually are) different, the matrix $\tau(i,j)$ is in general non-symmetric irrespective of the length of the time series.
Additionally, for each $i,j$ pair of ROIs we can obtain a set of delays for each individual source event $k$: $\tau^{(k)}(i,j)$, an average of these values $\bar{\tau}(i,j)$, or alternatively a delay between average events $\tau(i,j)$ (like the ones in Fig.~\ref{SelectEvents} D-F).

\section{Results:}
For the sake of presentation we will proceed here to discuss the performance of the method on two settings. The first (3.1) corresponds to the analysis of BOLD time series (n=90) from the AAL parcelation \cite{AAL} and the second (3.2) describes a voxel-wize functional connectivity analysis. In all cases we will compute classical Pearson correlation measures and compare them with ones obtained from the proposed method.

\subsection{Functional connectivity on AAL parcelled time series data sets}
The purpose of this subsection is to show how the computations explained previously perform on a single subject and a group of healthy participants, for this, we are using a group of 32 individuals from ABIDE preprocessed database~\citep{ABIDEdatabase}. The 90 AAL preprocessed (using Data Processing Assistant for Resting-State fMRI (DPARSF) pipeline) time series have been downloaded from the database and z-scored before lineal and nldFC computations.

\begin{figure} [h!]
\centering
\includegraphics[width=.60\textwidth]{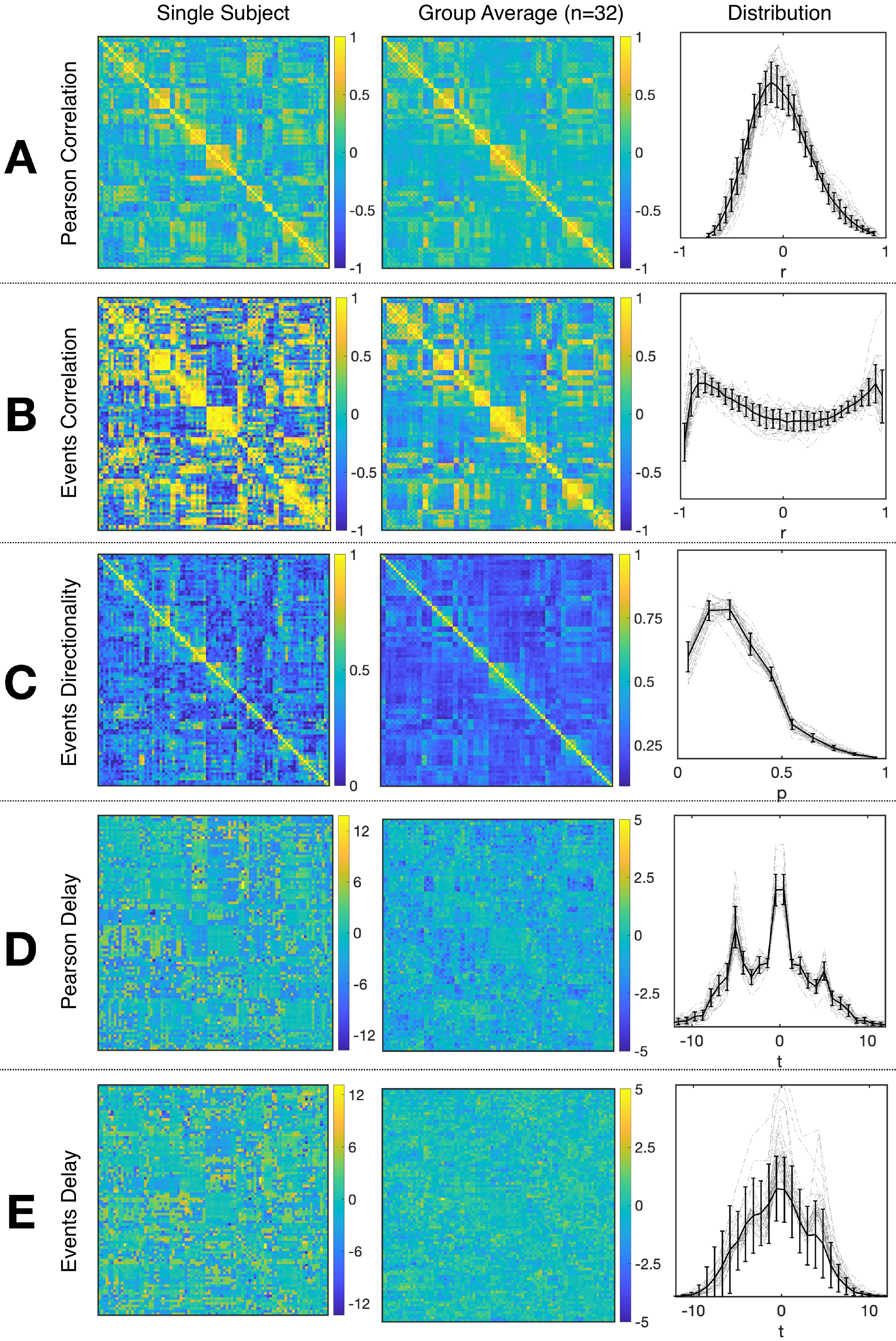}
\caption{Matrices and distributions of each calculation performed. First column are matrices (90 AAL atlas region time series) for a single subject, middle column shows the average matrix across subjects in the sample (n=79), and right column shows the distribution of each computation, where grey dashed lines are histograms for each subject, continuous black line are the average across subjects (errorbars are S.D.). Panel A)  Pearson correlation of the entire time series; Panel B) Correlation using only the large events; C) Directionality of the events. Note that the histogram shows that most of the regions share between .2 and .3 of events; D) Delays computed from whole time series Pearson correlation and E) Delays from large amplitude events.}
\label{ComputationsComparisons}
\end{figure}

The results of these computations can be seen in Figure~\ref{ComputationsComparisons}, which shows a matrix for single subject, a mean matrix of the whole group and the distribution of each computation. First, Fig.~\ref{ComputationsComparisons} A shows typical results obtained from Pearson correlations between all the time series, note that the distribution is a usual Gaussian shape. The visibly different shape of the histogram of event correlations as compared to the distribution of Pearson correlations in Fig.~\ref{ComputationsComparisons} B is simply due to the shape of the sampling distribution of Pearson estimator for small length of time series. This issue can be adjusted as discussed in Supplementary Material.
Fig.~\ref{ComputationsComparisons} C shows the distribution of asymmetry computed from events, here we are computing the proportion of shared events between regions, to see the first computation explained in methods section, subtracting the transposed matrix for each subject can be seen in Supplementary Material.
Delay between time series are showed in Fig.~\ref{ComputationsComparisons} D, for shifted time series as in \cite{Mitra2018} and Fig.~\ref{ComputationsComparisons} E, for delay computed using events. Note that for Fig.~\ref{ComputationsComparisons} D, the apparent asymmetry is due to the TRs subtracted at the beginning and end of the signal, to allow the computation, while for Fig.~\ref{ComputationsComparisons} E, the event selection between target and source, so it is not an artifact of the computation. 


\subsection{Replication of voxel-wise functional connectivity findings }
As a further validation of the computations explained above, we have
replicated~\cite{Xu2018} findings on functional connectivity between insular sub-regions on Autism Syndrome patients, using data from the ABIDE preprocessed database~\citep{ABIDEdatabase}. This is an open database with thousands of scans already preprocessed of Autistic patients (AU) and age-matched Healthy subjects (HS) \sloppy\url{http://preprocessed- connectomes-project.org/abide/quality_assessment.html} ~\citep{Rolls2016,Zheng2016,Dadi2019}. For this computations, we have used UCLA database from ABIDE, as in~\cite{Xu2018}, only subjects with good quality of the scan and under 1mm of frame-wise displacement were included into the sample, resulting a sample of 47 AU and 32 HS. The MRI data acquisition as the preprocessing pipeline used for this database can be accessed here: \url{http://preprocessed-connectomes-project.org/abide/Pipelines.html}.

\subsubsection{Pearson correlations}
For each subject scan, mean time series from six insular subregions (using Brainnetome functional atlas,~\cite{Brainnetome}) were extracted and correlated using Pearson correlation with all the rest of tha voxels of brain (grey matter masked) as in~\cite{Xu2018}. One-sample t-test was computed for each group of participants (AU and HS) to result the correlation pattern of each insular subregion, obtaining comparable results as in [\cite{Xu2018} (Fig.~\ref{SelectEvents})]. For purposes of space we only show results from left ventral agranular Insula subregion. As in~\cite{Xu2018}, HS resulted higher correlation of this ROI with bilateral precuneus cortex (see Figure~\ref{GroupComparisons1}). 

\subsubsection{non-linear functional co-activations}
\begin{figure} [h!]
 \centering
 \includegraphics[width=\textwidth]{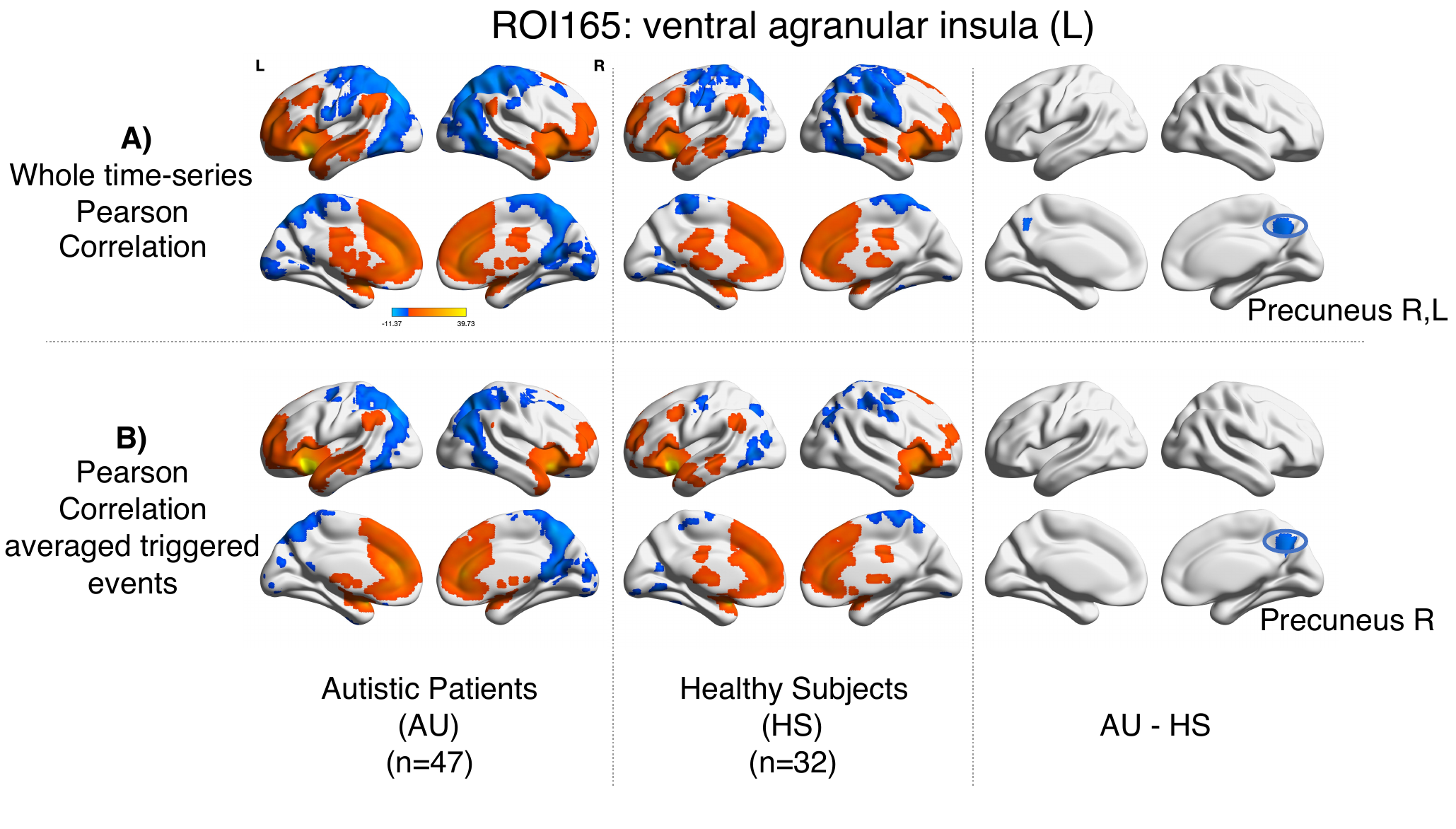}
 \caption{Comparison between Pearson correlation and the correlation of the large events.  Panel A) Results for zero-lag seed-voxel correlations using the left ventral agranular insula as seed. It is shown the Z-transformed Pearson correlation one-sample t-test for each group of participants (first-second columns for AU and third-fourth for HS) and two-sample t-test to assess differences between groups (fifth and sixth columns of brain surfaces)(GFR corrected voxel p<.001, cluster p<.05); Panel B) The same distribution of columns as A), for the results of  correlating  the the large events (Z-transformed, GFR corrected voxel p<.001, cluster p<.05);
 }
 \label{GroupComparisons1}
 \end{figure}

Following the method explained in~\cite{Tagliazucchi2011}, relevant events from mean time series of left ventral agranular insula subregion were extracted (triggering events where the amplitude is above a 1sd threshold, 2 TR previous to this trigger and 4 TR after). All time series from all the voxels in the brain (grey matter masking) corresponding to those events were extracted. Then correlations between the 'mean event' of insula and the 'mean event' of each voxel were computed. As it can be observed in Figure~\ref{GroupComparisons1} panel B, similar results to Pearson correlation of the whole signal  were obtained, it is important to note that here we are only taken into account the signal from events, not the whole time series. Computing a two-sample t-test (GFR corrected, p-voxel=.001, p-cluster=.05) the same cluster of higher correlation between insula and precuneus cortex in HS group can be observed.

\subsubsection{Asymmetry}
\begin{figure} [h!]
 \centering
 \includegraphics[width=\textwidth]{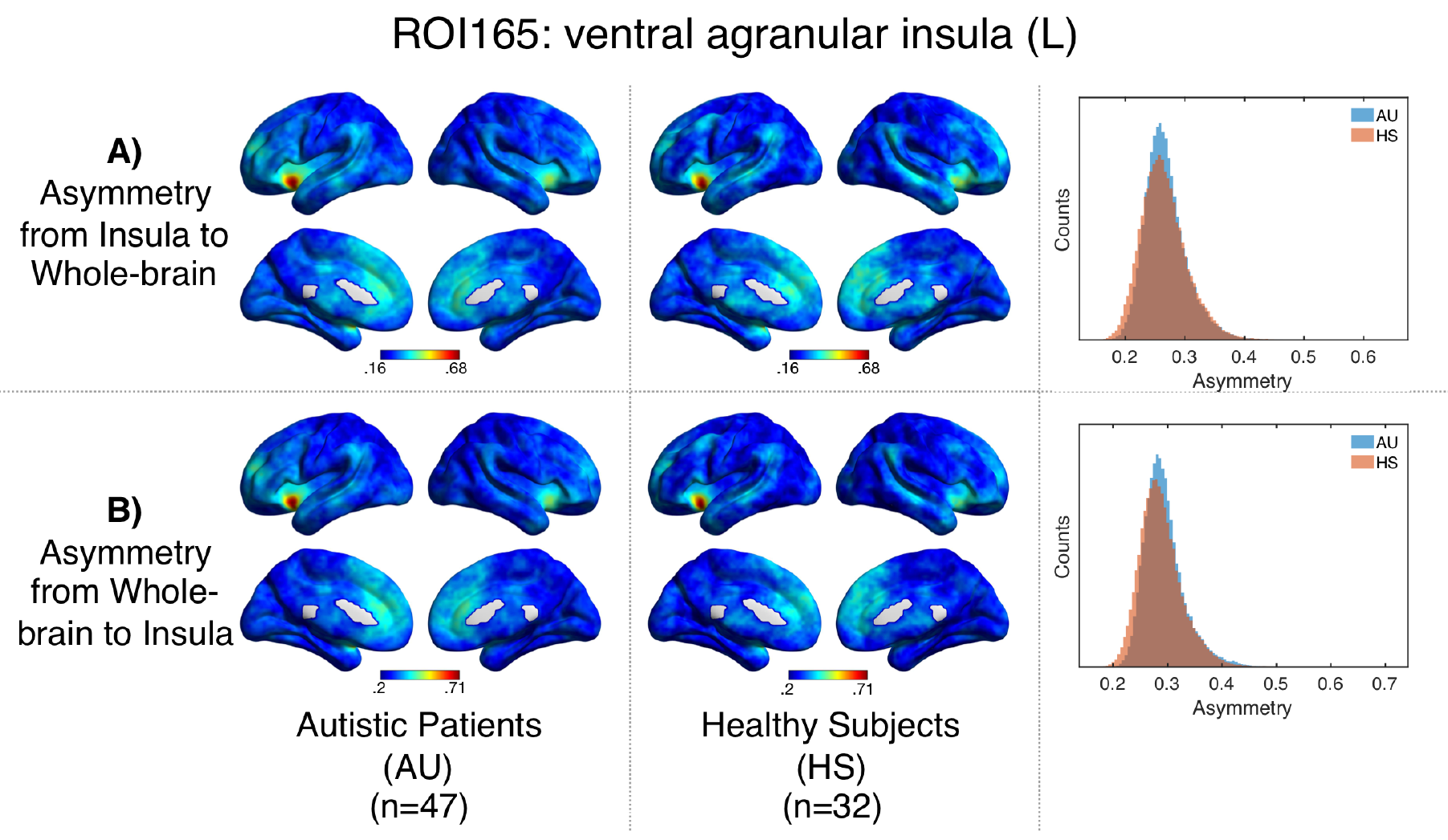}
 \caption{Directionality computed from-Insula and to-Insula for all brain voxels. Panel A shows group averages for asymmetry from Insula subregion mean time series to the rest of the brain for Autistic patients (left columns), Healthy subjects (middle columns) and histrograms showing the distribution of these directionalities. Panel B shows the same computations but for directionality from all the brain to the Insula subregion mean time series. 
 }
 \label{GroupComparisons2}
 \end{figure}

As it has been explained above, the correlation value between two signals (i,j) obtained when computing relevant events is not symmetric, the correlation of the events of a source with its target $r(i,j)$ is not necessarily the same than the correlation of the events of that target, acting as source, with the source, acting as a target $r(j,i)$. The difference between this $r(i,j)$ and $r(j,i)$ can be understood in terms of directionality of the correlations.
To test this and reveal new results on the activity of left ventral agranular insula, we computed these differences across the whole brain, resulting higher asymmetry of HS group between the ROI and Right superior frontal gyrus, and higher asymmetry of AU group between the ROI and right fusiform gyrus (Figure~\ref{GroupComparisons2}). 
 
\subsubsection{Delay}
\begin{figure} [h!]
 \centering
 \includegraphics[width=.8\textwidth]{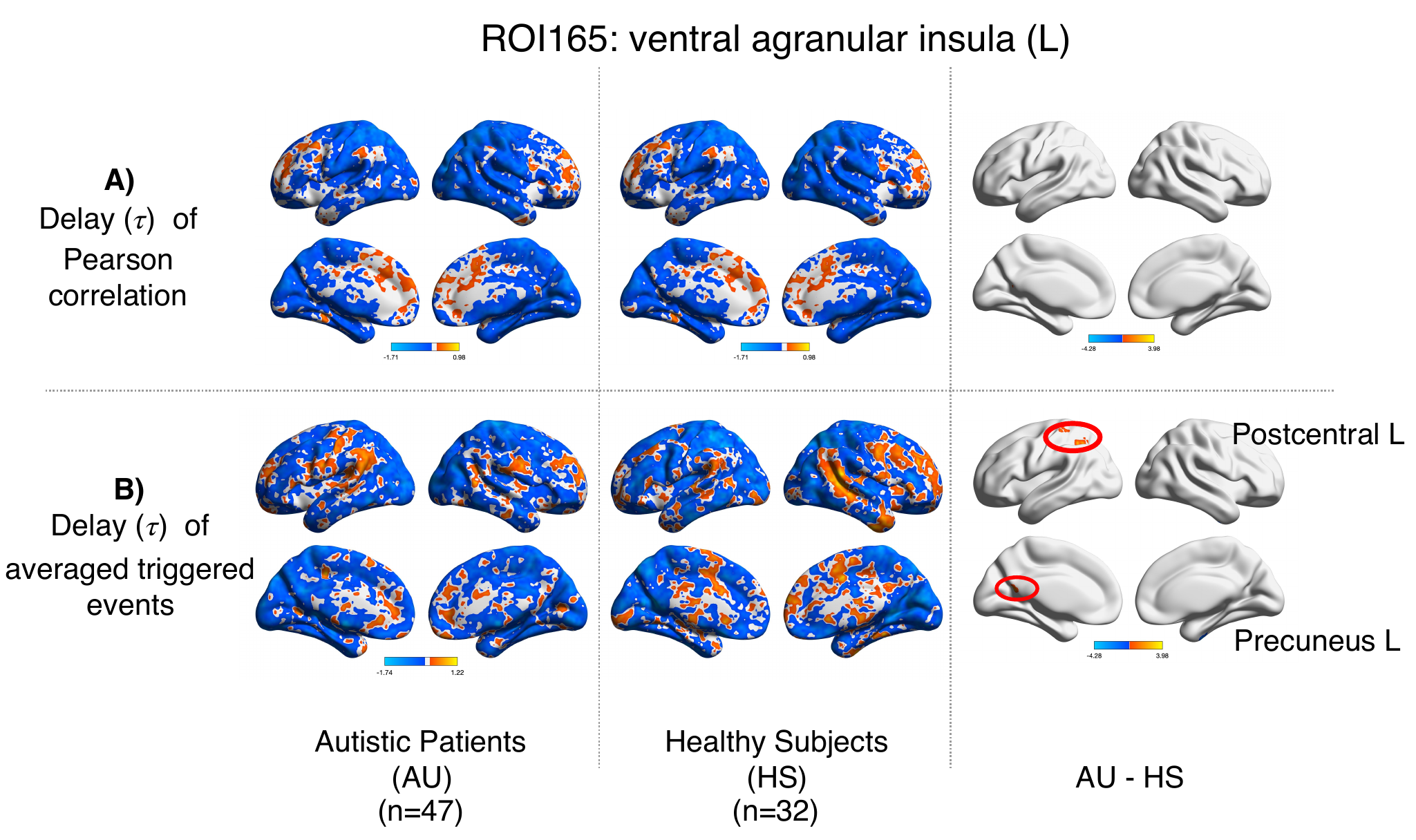}
 \caption{
 Delay computations for Autistic Syndrome patients  and Healthy subjects. Left columns show mean delays for Autistic patients, middle columns mean delays for Healthy subjects and right columns comparison between groups. Panel A) Delay ($\tau$) computed from Pearson correlations. Panel B) Delay ($\tau$) computed from averaged triggered events. Notice that the delay maps show differences which are not uncovered by Pearson correlation map or Pearson correlation delays computations (Figure~\ref{GroupComparisons1} panel A), as it can be seen in Postcentral gyrus difference in delay.
 }
 \label{GroupComparisons3}
 \end{figure}

\begin{figure} [h!]
 \centering
 \includegraphics[width=0.7  \textwidth]{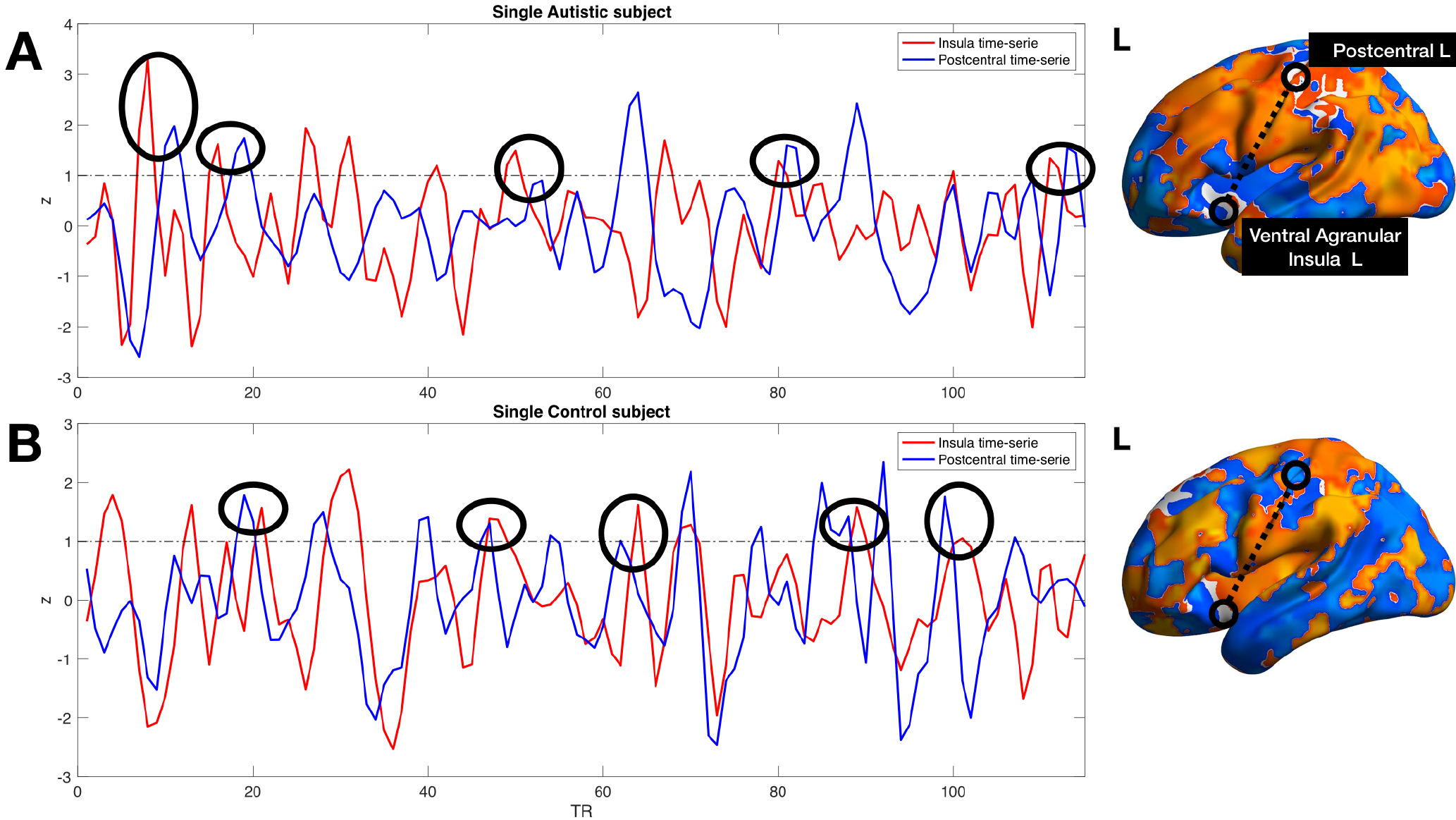}
 \caption{Single subject Postcentral gyrus (blue lines) and ventral agranular Insula (red lines) time series. Panel A shows example data from an Autism Syndrome patient, Panel B shows a healthy subject. Black circles show delays between those time series that support the group results obtained. Note that insular events on an AU patient are usually temporally before postcentral events, resulting on a negative INS-PC delay, while opposite results can be found on HS.  
 }
 \label{DelaySingle}
 \end{figure}

Al previous computations can be understood as lag-0 results, but it can also be computed the delay from an event, in a source time series, and the closest event in a target time series. To test this delays from left ventral agranular insula to all the rest of brain voxels were computed. Comparing the delays between groups it is shown that while left postcentral gyrus and the above mentioned precuneus cortex, have a positive delay for AU, while they have a negative delay for HS (Figure~\ref{GroupComparisons3}). To better explain the differences in delay $\tau$ between AU and HS subjects, Fig.\ref{DelaySingle} shows time series of postcentral gyrus and ventral agranula Insula for a single AU subject (Panel A) and a HS subject (Panel B).


\section{Discussion: Features, advantages and limitations of the proposed strategy} 

Emphasizing the relevance of relatively high amplitude BOLD signal while compressing the data motivated the two paradigms we have proposed previously; namely the point process~\cite{Tagliazucchi2012} and the rBeta technique~\cite{Tagliazucchi2011}. Both attempt to capture the spatio-temporal dynamics with the smallest possible sampling with a trade off between temporal and spatial resolution. The point process compresses in the temporal domain, which implies that to smoothly represent spatio-temporal correlated patterns one needs to sample more voxels. On the other hand, the rBeta approach uses much fewer voxels, but at the expense of keeping additional temporal information around each threshold crossing. These two variants  have demonstrated to main advantages comparing to the above mentioned functional connectivity measures, the fist one is that they imply a data size reduction and less computational resources to obtain comparable results to full time series analyses, and second, as these are only focusing on relevant high amplitude time-points, or events, less significant events occurring during the scan are not blurring the computations.

The non-linear dynamic functional connectivity method we are proposing, offers a widely different perspective in the analysis of brain co-activation patterns without much numerical complications, since it implies no more than thresholding and linear correlations, facilitating a simple interpretation of the resulting functional connectivity paths. The fact that the correlations are computed from events identified either as seeds or targets leads to directed graphs (i.e., asymmetric correlation matrices). These seed-target relations can lead to new approaches to understand brain dynamics, for instance, as in the example of Autism Syndrome in which  the computation of delays between events shows distinct information. Pearson correlation, computed between Left ventral agranular Insula and Postcentral gyrus, does not show any differences between Autistic patients and Healthy subjects, while it has been reported that Poscentral gyrus has a differential connectivity in Autism Syndrome when analyzing big samples~\cite{Gu2015}. When computing delays from Insula to other regions, differences between healthy subjects and Autism Syndrome in Postcentral gyrus appear, which leads us to think that this differential connectivity may not be just spatial but on a spatio-temporal domain. Another example is the additional information we have found on the comparison between Autism Syndrome and Healthy Subjects correlations which shows that Autism Syndrome  patients present a lower connectivity between precuneus cortex and ventral agranular insula, this explanation can be found on the differences in delay between these two areas (Fig. \ref{GroupComparisons3} panel B shows there is a higher delay on Autism Syndrome  from insula to precuneus, explaining the drop on correlation).

The main advantages provided by this new method of analysis are two: A) The results are highly reproducible on correlations asymmetry and delays, even changing the threshold used to extract the relevant events, that means that the method is not depending on specific cut-off point, as sometimes can happen when using $r-thresholding$ from Pearson correlations. B) This method is applicable to task data, for instance defining  the task timing (convoluted with HFR function) as the seed time series; or as an attempt to explain dynamic functional connectivity fluctuations, possibly due to on-going cognition, as suggested in \cite{Gonzalez_Castillo_2014}. 

Further testing of the method should be performed to identify more specifically the limitations of the method. A) For instance, we have not compared the method with results obtained from sliding-window Pearson correlation, a widely-used method to inspect dynamics in functional connectivity~\cite{Hutchison2013,Preti_2017}. In further work we expect that there will be a relation between this window based connectivity and the information provided from the delays of our method. B) Another technical issue that we are aware is the the few points we are correlating when selecting relevant events, which is the same limitation of sliding-windows correlations for dynamic connectivity analyses.  
C) It is unclear the meaning of  the delays distribution, something already intriguing from previous results obtained using  Pearson correlation delays (\cite{Mitra2015}, Fig.5).


Finally we shall mention that while here we concentrate on the activation events, i.e., denoted by the BOLD signal upward crossing of a threshold, the same method can be applied without modification to \emph {de-activation} events. In such a way graphs of regions of interest which are correlated with the deactivation of regions can be obtained, something that we are not aware was considered before.

In conclusion, we have analyzed undisclosed properties of the previously published rBeta method~\cite{Tagliazucchi2011}. Overall these calculations provide different kind of information, than the usual Pearson correlation of the entire time series. As a prof of concept  we have replicated a recently published study of  Functional Connectivity in  Autism Syndrome, adding the new features obtained from our method. More work has to be done to override the possible limitations of the method and to test it with task paradigms...

\section*{Conflict of Interest Statement}
The authors declare that the research was conducted in the absence of any commercial or financial relationships that could be construed as a conflict of interest.

\section*{Funding}
Work supported by the MICINN (Spain) grant PSI2017-82397-R, the National Science Centre (Poland) grant DEC-2015/17/D/ST2/03492, and by CONICET (Argentina) and Escuela de Ciencia y Tecnología, UNSAM.
Work conducted under the auspice of the Jagiellonian University-UNSAM Cooperation Agreement.

\section*{Data Availability Statement}
The MATLAB code accompanying the paper is available at \url{https://github.com/remolek/NFC}.

\bibliographystyle{frontiersinSCNS_ENG_HUMS} 

\begin{thebibliography}{33}
\providecommand{\natexlab}[1]{#1}
\expandafter\ifx\csname urlstyle\endcsname\relax
  \providecommand{\doi}[1]{doi:\discretionary{}{}{}#1}\else
  \providecommand{\doi}{doi:\discretionary{}{}{}\begingroup
  \urlstyle{rm}\Url}\fi
\providecommand{\selectlanguage}[1]{\relax}
\providecommand{\bibAnnoteFile}[1]{%
  \IfFileExists{#1}{\begin{quotation}\noindent\textsc{Key:} #1\\
  \textsc{Annotation:}\ \input{#1}\end{quotation}}{}}
\providecommand{\bibAnnote}[2]{%
  \begin{quotation}\noindent\textsc{Key:} #1\\
  \textsc{Annotation:}\ #2\end{quotation}}

\bibitem[{Allan et~al.(2015)Allan, Francis, Caballero-Gaudes, Morris, Liddle,
  Liddle et~al.}]{Allan2015}
Allan, T.~W., Francis, S.~T., Caballero-Gaudes, C., Morris, P.~G., Liddle,
  E.~B., Liddle, P.~F., et~al. (2015).
\newblock {Functional Connectivity in MRI Is Driven by Spontaneous BOLD
  Events}.
\newblock \emph{PLoS One} 10, e0124577.
\newblock \doi{10.1371/journal.pone.0124577}
\input{Allan2015}

\bibitem[{Amico et~al.(2014)Amico, Gomez, {Di Perri}, Vanhaudenhuyse,
  Lesenfants, Boveroux et~al.}]{Amico2014}
Amico, E., Gomez, F., {Di Perri}, C., Vanhaudenhuyse, A., Lesenfants, D.,
  Boveroux, P., et~al. (2014).
\newblock {Posterior Cingulate Cortex-Related Co-Activation Patterns: A Resting
  State fMRI Study in Propofol-Induced Loss of Consciousness}.
\newblock \emph{PLoS One} 9, e100012.
\newblock \doi{10.1371/journal.pone.0100012}
\input{Amico2014}

\bibitem[{Chen et~al.(2015)Chen, Chang, Greicius, and Glover}]{Chen2015}
Chen, J.~E., Chang, C., Greicius, M.~D., and Glover, G.~H. (2015).
\newblock {Introducing co-activation pattern metrics to quantify spontaneous
  brain network dynamics}.
\newblock \emph{Neuroimage} 111, 476--488.
\newblock \doi{10.1016/J.NEUROIMAGE.2015.01.057}
\input{Chen2015}

\bibitem[{Cifre et~al.(2017)Cifre, Zarepour, Horovitz, Cannas, and
  Chialvo}]{cifre2017few}
Cifre, I., Zarepour, M., Horovitz, S.~G., Cannas, S., and Chialvo, D.~R.
  (2017).
\newblock On why a few points suffice to describe spatiotemporal large-scale
  brain dynamics.
\newblock \emph{arXiv preprint arXiv:1707.00759}
\input{cifre2017few}

\bibitem[{Craddock et~al.(2013)Craddock, Benhajali, Chu, Chouinard, Evans,
  Jakab et~al.}]{ABIDEdatabase}
Craddock, C., Benhajali, Y., Chu, C., Chouinard, F., Evans, A., Jakab, A.,
  et~al. (2013).
\newblock The neuro bureau preprocessing initiative: open sharing of
  preprocessed neuroimaging data and derivatives <br />.
\newblock \emph{Frontiers in Neuroinformatics}
  \doi{10.3389/conf.fninf.2013.09.00041}
\input{ABIDEdatabase}

\bibitem[{Dadi et~al.(2019)Dadi, Rahim, Abraham, Chyzhyk, Milham, Thirion
  et~al.}]{Dadi2019}
Dadi, K., Rahim, M., Abraham, A., Chyzhyk, D., Milham, M., Thirion, B., et~al.
  (2019).
\newblock {Benchmarking functional connectome-based predictive models for
  resting-state fMRI}.
\newblock \emph{NeuroImage} \doi{10.1016/j.neuroimage.2019.02.062}
\input{Dadi2019}

\bibitem[{Egu{\'{i}}luz et~al.(2005)Egu{\'{i}}luz, Chialvo, Cecchi, Baliki, and
  Apkarian}]{Eguiluz2005}
Egu{\'{i}}luz, V.~M., Chialvo, D.~R., Cecchi, G.~A., Baliki, M., and Apkarian,
  A.~V. (2005).
\newblock {Scale-Free Brain Functional Networks}.
\newblock \emph{Phys. Rev. Lett.} 94, 018102.
\newblock \doi{10.1103/PhysRevLett.94.018102}
\input{Eguiluz2005}

\bibitem[{Fan et~al.(2016)Fan, Chu, Li, Chen, Xie, Zhang et~al.}]{Brainnetome}
Fan, L., Chu, C., Li, H., Chen, L., Xie, S., Zhang, Y., et~al. (2016).
\newblock {The Human Brainnetome Atlas: A New Brain Atlas Based on Connectional
  Architecture}.
\newblock \emph{Cerebral Cortex} 26, 3508--3526.
\newblock \doi{10.1093/cercor/bhw157}
\input{Brainnetome}

\bibitem[{Finn et~al.(2015)Finn, Shen, Scheinost, Rosenberg, Huang, Chun
  et~al.}]{Finn2015}
Finn, E.~S., Shen, X., Scheinost, D., Rosenberg, M.~D., Huang, J., Chun, M.~M.,
  et~al. (2015).
\newblock {Functional connectome fingerprinting: Identifying individuals using
  patterns of brain connectivity}.
\newblock \emph{Nature Neuroscience} 18, 1664--1671.
\newblock \doi{10.1038/nn.4135}
\input{Finn2015}

\bibitem[{Gonzalez-Castillo et~al.(2014)Gonzalez-Castillo, Handwerker,
  Robinson, Hoy, Buchanan, Saad et~al.}]{Gonzalez_Castillo_2014}
Gonzalez-Castillo, J., Handwerker, D.~A., Robinson, M.~E., Hoy, C.~W.,
  Buchanan, L.~C., Saad, Z.~S., et~al. (2014).
\newblock {The spatial structure of resting state connectivity stability on the
  scale of minutes}.
\newblock \emph{Front. Neurosci.} 8.
\newblock \doi{10.3389/fnins.2014.00138}
\input{Gonzalez_Castillo_2014}

\bibitem[{Gu et~al.(2015)Gu, Zhang, Rolls, Feng, and Cheng}]{Gu2015}
Gu, H., Zhang, J., Rolls, E.~T., Feng, J., and Cheng, W. (2015).
\newblock {Autism: reduced connectivity between cortical areas involved in face
  expression, theory of mind, and the sense of self}.
\newblock \emph{Brain} 138, 1382--1393.
\newblock \doi{10.1093/brain/awv051}
\input{Gu2015}

\bibitem[{Hutchison et~al.(2013)Hutchison, Womelsdorf, Allen, Bandettini,
  Calhoun, Corbetta et~al.}]{Hutchison2013}
Hutchison, R.~M., Womelsdorf, T., Allen, E.~A., Bandettini, P.~A., Calhoun,
  V.~D., Corbetta, M., et~al. (2013).
\newblock {Dynamic functional connectivity: Promise, issues, and
  interpretations}.
\newblock \emph{NeuroImage} 80, 360--378.
\newblock \doi{10.1016/j.neuroimage.2013.05.079}
\input{Hutchison2013}

\bibitem[{Jiang et~al.(2014)Jiang, Lv, Zhu, Zhang, Hu, Guo et~al.}]{Jiang2014}
Jiang, X., Lv, J., Zhu, D., Zhang, T., Hu, X., Guo, L., et~al. (2014).
\newblock {Integrating group-wise functional brain activities via point
  processes}.
\newblock In \emph{2014 IEEE 11th Int. Symp. Biomed. Imaging} (IEEE), 669--672.
\newblock \doi{10.1109/ISBI.2014.6867959}
\input{Jiang2014}

\bibitem[{Li et~al.(2014)Li, Li, Hu, Chen, and Dai}]{Li2014}
Li, W., Li, Y., Hu, C., Chen, X., and Dai, H. (2014).
\newblock {Point process analysis in brain networks of patients with diabetes}.
\newblock \emph{Neurocomputing} 145, 182--189.
\newblock \doi{10.1016/J.NEUCOM.2014.05.045}
\input{Li2014}

\bibitem[{Liu et~al.(2013)Liu, Chang, and Duyn}]{Liu2013}
Liu, X., Chang, C., and Duyn, J.~H. (2013).
\newblock {Decomposition of spontaneous brain activity into distinct fMRI
  co-activation patterns}.
\newblock \emph{Front. Syst. Neurosci.} 7, 101.
\newblock \doi{10.3389/fnsys.2013.00101}
\input{Liu2013}

\bibitem[{Liu and Duyn(2013)}]{Liu2013b}
Liu, X. and Duyn, J.~H. (2013).
\newblock {Time-varying functional network information extracted from brief
  instances of spontaneous brain activity}.
\newblock \emph{Proc. Natl. Acad. Sci.} 110, 4392--4397.
\newblock \doi{10.1073/pnas.1216856110}
\input{Liu2013b}

\bibitem[{Mitra and Raichle(2016)}]{Mitra2016}
Mitra, A. and Raichle, M.~E. (2016).
\newblock {How networks communicate: propagation patterns in spontaneous brain
  activity}.
\newblock \emph{Philos. Trans. R. Soc. B Biol. Sci.} 371, 20150546.
\newblock \doi{10.1098/rstb.2015.0546}
\input{Mitra2016}

\bibitem[{Mitra and Raichle(2018)}]{Mitra2018}
Mitra, A. and Raichle, M.~E. (2018).
\newblock {Principles of cross-network communication in human resting state
  fMRI}.
\newblock \emph{Scand. J. Psychol.} 59, 83--90.
\newblock \doi{10.1111/sjop.12422}
\input{Mitra2018}

\bibitem[{Mitra et~al.(2015{\natexlab{a}})Mitra, Snyder, Blazey, and
  Raichle}]{Mitra2015}
Mitra, A., Snyder, A.~Z., Blazey, T., and Raichle, M.~E. (2015{\natexlab{a}}).
\newblock {Lag threads organize the brain's intrinsic activity}.
\newblock \emph{Proc. Natl. Acad. Sci.} 112, E2235--E2244.
\newblock \doi{10.1073/PNAS.1503960112}
\input{Mitra2015}

\bibitem[{Mitra et~al.(2015{\natexlab{b}})Mitra, Snyder, Constantino, and
  Raichle}]{Mitra2015a}
Mitra, A., Snyder, A.~Z., Constantino, J.~N., and Raichle, M.~E.
  (2015{\natexlab{b}}).
\newblock {The Lag Structure of Intrinsic Activity is Focally Altered in High
  Functioning Adults with Autism}.
\newblock \emph{Cereb. Cortex} 27.
\newblock \doi{10.1093/cercor/bhv294}
\input{Mitra2015a}

\bibitem[{Mitra et~al.(2014)Mitra, Snyder, Hacker, and Raichle}]{Mitra2014}
Mitra, A., Snyder, A.~Z., Hacker, C.~D., and Raichle, M.~E. (2014).
\newblock {Lag structure in resting-state fMRI}.
\newblock \emph{J. Neurophysiol.} 111, 2374--2391.
\newblock \doi{10.1152/jn.00804.2013}
\input{Mitra2014}

\bibitem[{Ochab et~al.(2019)Ochab, Tarnowski, Nowak, and Chialvo}]{Ochab2019}
Ochab, J.~K., Tarnowski, W., Nowak, M.~A., and Chialvo, D.~R. (2019).
\newblock {On the pros and cons of using temporal derivatives to assess brain
  functional connectivity}.
\newblock \emph{Neuroimage} 184, 577--585.
\newblock \doi{10.1016/J.NEUROIMAGE.2018.09.063}
\input{Ochab2019}

\bibitem[{Petridou et~al.(2013)Petridou, Gaudes, Dryden, Francis, and
  Gowland}]{Petridou2013}
Petridou, N., Gaudes, C.~C., Dryden, I.~L., Francis, S.~T., and Gowland, P.~A.
  (2013).
\newblock {Periods of rest in fMRI contain individual spontaneous events which
  are related to slowly fluctuating spontaneous activity}.
\newblock \emph{Hum. Brain Mapp.} 34, 1319--1329.
\newblock \doi{10.1002/hbm.21513}
\input{Petridou2013}

\bibitem[{Preti et~al.(2017)Preti, Bolton, and {Van De Ville}}]{Preti_2017}
Preti, M.~G., Bolton, T.~A., and {Van De Ville}, D. (2017).
\newblock {The dynamic functional connectome: State-of-the-art and
  perspectives}.
\newblock \emph{Neuroimage} 160, 41--54.
\newblock \doi{10.1016/j.neuroimage.2016.12.061}
\input{Preti_2017}

\bibitem[{Rolls et~al.(2016)Rolls, Zhang, Feng, Wan, Cheng, Luo
  et~al.}]{Rolls2016}
Rolls, E.~T., Zhang, J., Feng, J., Wan, L., Cheng, W., Luo, Q., et~al. (2016).
\newblock {Functional connectivity decreases in autism in emotion, self, and
  face circuits identified by Knowledge-based Enrichment Analysis}.
\newblock \emph{NeuroImage} 148, 169--178.
\newblock \doi{10.1016/j.neuroimage.2016.12.068}
\input{Rolls2016}

\bibitem[{Tagliazucchi et~al.(2012)Tagliazucchi, Balenzuela, Fraiman, and
  Chialvo}]{Tagliazucchi2012}
Tagliazucchi, E., Balenzuela, P., Fraiman, D., and Chialvo, D.~R. (2012).
\newblock {Criticality in Large-Scale Brain fMRI Dynamics Unveiled by a Novel
  Point Process Analysis}.
\newblock \emph{Front. Physiol.} 3, 15.
\newblock \doi{10.3389/fphys.2012.00015}
\input{Tagliazucchi2012}

\bibitem[{Tagliazucchi et~al.(2011)Tagliazucchi, Balenzuela, Fraiman, Montoya,
  and Chialvo}]{Tagliazucchi2011}
Tagliazucchi, E., Balenzuela, P., Fraiman, D., Montoya, P., and Chialvo, D.~R.
  (2011).
\newblock {Spontaneous BOLD event triggered averages for estimating functional
  connectivity at resting state}.
\newblock \emph{Neurosci. Lett.} 488, 158--163.
\newblock \doi{10.1016/j.neulet.2010.11.020}
\input{Tagliazucchi2011}

\bibitem[{Tagliazucchi et~al.(2016)Tagliazucchi, Siniatchkin, Laufs, and
  Chialvo}]{Tagliazucchi2016}
Tagliazucchi, E., Siniatchkin, M., Laufs, H., and Chialvo, D.~R. (2016).
\newblock {The Voxel-Wise Functional Connectome Can Be Efficiently Derived from
  Co-activations in a Sparse Spatio-Temporal Point-Process}.
\newblock \emph{Front. Neurosci.} 10, 381.
\newblock \doi{10.3389/fnins.2016.00381}
\input{Tagliazucchi2016}

\bibitem[{Tzourio-Mazoyer et~al.(2002)Tzourio-Mazoyer, Landeau, Papathanassiou,
  Crivello, Etard, Delcroix et~al.}]{AAL}
Tzourio-Mazoyer, N., Landeau, B., Papathanassiou, D., Crivello, F., Etard, O.,
  Delcroix, N., et~al. (2002).
\newblock {Automated Anatomical Labeling of Activations in SPM Using a
  Macroscopic Anatomical Parcellation of the MNI MRI Single-Subject Brain}.
\newblock \emph{Neuroimage} 15, 273--289.
\newblock \doi{10.1006/NIMG.2001.0978}
\input{AAL}

\bibitem[{van~den Heuvel and {Hulshoff P.}(2010)}]{VandenHeuvel2010}
van~den Heuvel, M.~P. and {Hulshoff P.}, H.~E. (2010).
\newblock {Exploring the brain network: A review on resting-state fMRI
  functional connectivity}.
\newblock \emph{European Neuropsychopharmacology} 20, 519--534.
\newblock \doi{DOI: 10.1016/j.euroneuro.2010.03.008}
\input{VandenHeuvel2010}

\bibitem[{Wu et~al.(2013)Wu, Liao, Stramaglia, Ding, Chen, and
  Marinazzo}]{Wu2013}
Wu, G.-R., Liao, W., Stramaglia, S., Ding, J.-R., Chen, H., and Marinazzo, D.
  (2013).
\newblock {A blind deconvolution approach to recover effective connectivity
  brain networks from resting state fMRI data}.
\newblock \emph{Med. Image Anal.} 17, 365--374.
\newblock \doi{10.1016/J.MEDIA.2013.01.003}
\input{Wu2013}

\bibitem[{Xu et~al.(2018)Xu, Wang, Zhang, Xu, Li, Zhou et~al.}]{Xu2018}
Xu, J., Wang, H., Zhang, L., Xu, Z., Li, T., Zhou, Z., et~al. (2018).
\newblock {Both Hypo-Connectivity and Hyper-Connectivity of the Insular
  Subregions Associated With Severity in Children With Autism Spectrum
  Disorders}.
\newblock \emph{Front. Neurosci.} 12, 234.
\newblock \doi{10.3389/fnins.2018.00234}
\input{Xu2018}

\bibitem[{Zheng et~al.(2016)Zheng, Xie, Chen, Yao, Zheng, Liu
  et~al.}]{Zheng2016}
Zheng, F., Xie, Y., Chen, X., Yao, Z., Zheng, W., Liu, G., et~al. (2016).
\newblock {Resting-State Time-Varying Analysis Reveals Aberrant Variations of
  Functional Connectivity in Autism}.
\newblock \emph{Frontiers in Human Neuroscience} 10, 1--11.
\newblock \doi{10.3389/fnhum.2016.00463}
\input{Zheng2016}

\end{thebibliography}

\end{document}